\documentstyle[twoside,fleqn,espcrc2]{article}

\def\calo{{\cal O}}

\def\ie{{\it i.e.}}

\def\be{\begin{equation}}
\def\ee{\end{equation}}
\def\bea{\begin{eqnarray}}
\def\eea{\end{eqnarray}}
\def\bmu{B_{\mu}} 

\def\pl#1#2#3{{\it Phys. Lett. }{\bf B#1~}(19#2)~#3}

\def\prl#1#2#3{{\it Phys. Rev. Lett. }{\bf #1~}(19#2)~#3}

\def\prep#1#2#3{{\it Phys. Rep. }{\bf #1~}(19#2)~#3}
\def\pr#1#2#3{{\it Phys. Rev. }{\bf D#1~}(19#2)~#3}
\def\np#1#2#3{{\it Nucl. Phys. }{\bf B#1~}(19#2)~#3}

\input epsf
\newcommand{\postscript}[2]
{\setlength{\epsfxsize}{#2\hsize}
\centerline{\epsfbox{#1}}}

\newcommand{\AmS}{{\protect\the\textfont2
  A\kern-.1667em\lower.5ex\hbox{M}\kern-.125emS}}

\title{Advantages and Disadvantages of Supersymmetry Breaking\\
          at Low Energies \thanks{Talk given at the  
4th International Conference on Supersymmetry (SUSY 96),
University of Maryland, May 1996.}}

\author{Alex Pomarol\address{Theory Division, CERN,\\
CH-1211 Geneva 23, Switzerland}}%

\begin{document}

\begin{abstract}
The breaking of supersymmetry is usually assumed to occur in a hidden sector.
Two natural candidates for the supersymmetry breaking transmission from
the hidden to the observable sector are gravity and the gauge interactions. 
Only the second one allows for supersymmetry breaking at low energies.
I show how the two candidates deal with the flavor and the $\mu$-problem;
I also briefly comment on the doublet-triplet and dark matter problem.
\end{abstract}

\maketitle

\section{INTRODUCTION}

The mass sum rule \cite{fer}, 
STr${\cal M}^2=0$, is a severe
constraint
in the search for realistic models based on supersymmetry.
To circumvent this {\it tree-level} mass relation,
one is  forced to assume that
the  breaking of supersymmetry originates in a  hidden sector 
that does not couple  {\it at tree level} to the observable sector. 

Two natural candidates for the supersymmetry breaking transmission from the 
hidden to the observable sector are the 
standard model (SM) gauge interactions and gravity. 
These interactions, being flavor-symmetric, induce 
universal squark masses avoiding 
 the supersymmetric flavor
problem (see next section). 

In theories with gravity mediating the supersymmetry
breaking (GravMSB) \cite{gra}, the  scalar and
 gaugino soft masses originate from 
 non-renormalizable operators induced by gravity:
\bea
&&\int d^4\theta 
\frac{XX^\dagger}{M^2_P}\phi\phi^\dagger \ \ \Longrightarrow\ \ 
m_0^2\sim\frac{F^2}{M^2_P}\, ,\label{opem0}\\
&&\int d^2\theta \frac{X}{M_P}WW\ \ \Longrightarrow\ \ 
m_{\lambda}\sim\frac{F}{M_P}\, ,\label{opemg}
\eea
where $\phi$ generically denotes a  superfield from the observable
sector while
 $X$ denotes
 a superfield  from the hidden sector whose $F$-component gets a VEV
and breaks supersymmetry.
Since the scalar masses $m_0$ are needed to be around the weak scale,
supersymmetry has to be broken at an intermediate scale
$\sqrt{F}\sim 10^{10}$ GeV.

In gauge mediated supersymmetry breaking (GMSB) theories
\cite{gaug}--\cite{us},
one introduces new  vector-like states ($\Phi$+$\bar\Phi$), called
 the ``messengers'', which 
 transform non-trivially under the SM gauge group.
These messengers couple directly to $X$ (the field that parametrizes
the supersymmetry breaking)
\be
W=\Phi\bar\Phi X\, ,
\ee
and induce
at the two- and one-loop level respectively 
   the operators (\ref{opem0}) and (\ref{opemg}) 
 [with
$M_P$ replaced by the messenger mass $M$]. 
Therefore, 
  soft masses are generated 
at a different  loop level
\be
m^2_0 \sim \left(\frac{\alpha}{4\pi}\right)^2
\frac{F^2}{M^2}\ ,
\ \ \ \ \
m_\lambda\sim \frac{\alpha}{4\pi}\frac{F}{M}\, .
\label{masses}
\ee
 Because of the
different dimensionalities between the fermionic and scalar mass terms, this
implies that the gaugino and squark masses  are of the same order,
$m_{\lambda}\sim { m_0}$.
The ratio $F/M$ must be in the range $10$--$100$ TeV to generate soft masses of
$\calo(M_Z)$. Thus,
this second scenario, unlike 
the GravMSB scenario,
 allows for supersymmetry breaking at low energies,
$\sqrt{F}\sim M\sim 10$ TeV.

Both possibilities, GravMSB and GMSB,  represent 
two compelling  scenarios of  low-energy supersymmetry.
In this talk, I will make a short ``tour'' around the 
different ways that  these two alternatives deal
with the different problems of supersymmetric theories.
In particular, I will focus on 
the supersymmetric flavor problem, the $\mu$-problem, and will mention
about the doublet-triplet problem and the dark matter problem.
For the different phenomenology
of these two scenarios
see the talks by M. Dine, G. Kane and S. Thomas at this meeting; 
see  also ref.~\cite{exp}.

\section{THE FLAVOR PROBLEM}

The experimental indication of small
flavor changing neutral currents (FCNC)
implies the need for
a super-GIM mechanism in the scalar-quark sector, \ie\
 the squarks of the first and second family  have to be   highly
degenerate.
This requirement is often called the supersymmetric flavor problem
\cite{hkr}.

In  GravMSB theories, the 
induced squark
 soft masses   are universal
(for a flat K\"ahler potential).
However, the scale at which these masses are induced is  the Planck scale, 
and
it is not  guaranteed that the degeneracy will still hold
at low energies. In fact, 
deviations from the universal values are usually too large 
in theories of flavor \cite{hkr,dp}; squark mass 
splittings can result  from the $M_P$--$M_Z$ running \cite{hkr}  or
from integrating out the heavy modes \cite{dp} \footnote{Unless one imposes 
non-Abelian family symmetries (see the talks by L. Hall and
Z. Berezhiani at this meeting).}.

In GMSB theories, the flavor problem is naturally solved
as gauge interactions provide flavor-symmetric
supersymmetry-breaking terms in the observable sector.
Moreover, in these theories,
since supersymmetry is broken at low energies,
 the soft masses are decoupled from any high energy sector and do not 
 suffer from the problem of GravMSB theories.

In short, breaking supersymmetry at
high (low) energies implies  that the scalar soft masses  feel (do not feel)  
the physics at   the ultraviolet  which can induce large FCNC.

\section{THE $\mu$-PROBLEM}

The $\mu$-problem is  
the difficulty in generating the correct supersymmetric 
mass  for the Higgs 
\be
W=\mu {\bar H}H\, ,
\label{mu}
\ee
which, for phenomenological reasons, has to be of the order of the weak
scale. A priori, one would expect this mass to be of
the Planck scale  or some other fundamental large mass
scales.

In GravMSB theories, the above puzzle can be solved in 
several ways. The most appealing solution \cite{giu} 
(at least to me) is to assume that
the term in eq.~(\ref{mu}) is forbidden in the limit 
of exact supersymmetry, and arises from  non-renormalizable operators
when supersymmetry is broken; alike the 
soft  terms (\ref{opem0}) and (\ref{opemg}). 
The 
effective  non-renormalizable operators
that generate a $\mu$ and a $\bmu$  (the soft scalar
mass $\bmu H\bar H$)
term are
\bea
&&\int d^4\theta 
\frac{X^\dagger}{M}H\bar H \ \ \Longrightarrow\ \ 
\mu\sim\frac{F}{M}\, ,\label{opemu}\\
&&\int d^4\theta 
\frac{XX^{\dagger}}{M^2}H\bar H \ \ \Longrightarrow\ \ 
\bmu\sim\frac{F^2}{M^2}\, ,\label{opebmu}
\eea
where $M$ has to be identified with $M_{P}$.

In theories of supersymmetry breaking at low energies, if
 we want to generate   $\mu$ and $\bmu$ from the 
operators in eqs.~(\ref{opemu})
and (\ref{opebmu}), 
these have  to be induced not at $M_{P}$ 
but at the messenger scale $M$ (since now the
supersymmetry breaking scale is much smaller $\sqrt{F}\sim$ 10 TeV).
Furthermore, since the  soft masses (\ref{masses})
are suppressed by  loop factors
with respect to $F/M$, we need  the operator  (\ref{opemu})  to be
generated at one loop while the operator
(\ref{opebmu}), being of dimension mass-squared,   
to be generated at two loops. 

The
operators (\ref{opemu})
and (\ref{opebmu}) break a Peccei-Quinn symmetry and cannot be induced
by gauge interactions alone, as the other soft masses (\ref{masses}).
Thus, we have to introduce new interactions in the model.
The simplest possibility is
to couple the Higgs superfields directly
to the messengers:
\be
W=\lambda H\Phi_1\Phi_2+\bar\lambda \bar H\bar\Phi_1\bar\Phi_2\, .
\label{couplinga}
\ee
Thus,
the operator (\ref{opemu}) is generated at one loop from
the diagram of fig.~1a and induces a  $\mu$ parameter 
of the right order
\be
\mu\sim\frac{\lambda\bar\lambda}{16\pi^2}\frac{F}{M}\, .
\label{muloop}
\ee
(For the exact result see ref.~\cite{us}.)
However with an extra insertion of the spurion
superfield $X$ in the messenger loop (diagram of fig.~1b), the
 operator (\ref{opebmu}) is also generated;  
thus $\bmu$ arises at one loop:
\be
\bmu\sim\frac{\lambda\bar\lambda}{16\pi^2}\left(\frac{F}{M}\right)^2
\, .\label{bmuloop}
\ee
{}From eqs.~(\ref{muloop}) and 
(\ref{bmuloop}) we obtain
\be
\bmu\sim\mu\frac{F}{M}\, .\label{problematic}
\ee
This problematic relation is the expression of the $\mu$-problem in
GMSB theories \cite{us}. It is just a consequence of generating both $\mu$
and $\bmu$ at the one-loop level through the same interactions.
Recall that $F/M\sim 10$--100 TeV,  and then 
eq.~(\ref{problematic}) implies that 
either $\mu$ is at the weak scale and $B_\mu$ violates
 naturalness, or $B_\mu$ is at the weak scale
and $\mu$ is unacceptably small.

\begin{figure}[htb]
\vspace{9pt}
\postscript{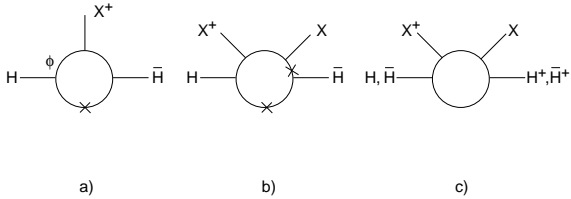}{1}
\caption{Superfield Feynman diagrams for generating
one-loop contributions to (a)
$\mu$, (b) $\bmu$, and (c) $m_H^2$, $m_{\bar H}^2$.
The internal lines with (without) a ``$\times$''  denote
a messenger $\langle\Phi\Phi\rangle$ ($\langle\Phi^\dagger\Phi\rangle$)
propagator.}
\label{fig1}
\end{figure}

Another 
possibility proposed in the literature to solve
the $\mu$-problem is to add an extra light Higgs superfield
$S$  with a 
superpotential 
\be
W=\lambda H\bar HS+\lambda'S^3\, .
\ee
Thus, $\mu$ and $\bmu$ are generated whenever $\langle S\rangle$
and $\langle F_S\rangle$ are non-zero.
This option is perfectly viable in GravMSB theories \cite{white}.
However, in GMSB theories 
trilinears, bilinears and the soft-mass of $S$ are suppressed with 
respect to the $H$ and $\bar H$ soft masses, and 
a non-zero VEV for $S$ requires an appreciable fine-tuning \cite{din}.
This problem can be overcome but its solution may require  
additional quark superfields
coupled to 
$S$ in order to induce a large soft mass $m_S^2$ \cite{din}
(for a recent idea see ref.~\cite{dns}).

\subsection{A Natural Solution to the $\mu$-Problem in GMSB Theories}

I will describe here a mechanism that satisfies the criteria
of naturalness \cite{us}:
{\it i)} the different 
Higgs parameters are generated by a single mechanism; {\it ii)} $\mu$
is generated at one loop, while $B_\mu$, $m_H^2$, $m_{\bar H}^2$ are
generated at two loops; {\it iii)} all new coupling constants 
are of order one;
{\it iv)} there are no new particles at the weak scale.
The idea behind this mechanism is to generate
the $\mu$ parameter, not from
the operator (\ref{opemu}), 
 but from the operator
\be
\int d^4\theta 
\frac{D^2\left[X^\dagger X\right]}{M^4}H\bar H \, .
\label{qqq}
\ee
Here $D_\alpha$ is the supersymmetric covariant derivative. 
This operator can be generated at one-loop level
from the diagram of fig.~2. 
To see that, one can  proceed in two steps. First, one can see that 
the loop just induces the
 operator $\int d^4\theta XX^\dagger S^\dagger$.
Secondly, one can integrate out the heavy singlet $S$ at tree level using
its equation of motion \cite{graphs}.
The crucial point about the diagram of fig.~2
is  that a $\bmu$-term cannot be induced 
from such a diagram even if we added
extra $X$ and $X^\dagger$ insertions in the loop. This is because a 
$D^2$ acting on any function of $X$ and $X^\dagger$ always produces
an antichiral superfield. 

This mechanism requires at least
two singlets, $S$ and 
$N$, such that only $S$  couples at tree level to 
$H\bar H$ and to the messengers.
A term $S^2$ is forbidden in the superpotential
to guarantee that the operator (\ref{opebmu}) is not generated.
An explicit model with the above requirements is given by
the superpotential
\be
W=S\big(\lambda_1 H \bar H +\frac{\lambda_2}{2}N^2+\lambda \Phi \bar \Phi
-M_N^2\big)~.
\label{spot}
\ee
The terms in eq.~(\ref{spot}) can be guaranteed by 
a discrete parity of the superfield $N$ and an R-symmetry. We believe
that eq.~(\ref{spot}) describes the simplest example in which the above
mechanism is operative. 
Notice that we have introduced a new mass parameter $M_N$ in eq.~(\ref{spot});
it is assumed to have the same origin as 
 the other scales in the model, $\sqrt{F}$ and $M$.

\begin{figure}[htb]
\postscript{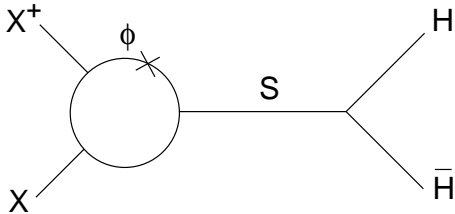}{.8}
\caption{Superfield Feynman diagram for generating a
one-loop contribution to
$\mu$. Same notation as in fig.~1. The internal line with an $S$ denotes
a  $\langle S^\dagger S\rangle$ propagator.}
\label{fig2}
\end{figure}

Including  the messenger one-loop corrections and minimizing the potential,
one finds \cite{us} that  $N$ gets a VEV of order  $M_N$ and becomes heavy 
together with $S$; only the Higgs doublets
$H$ and $\bar H$ remain light.
Integrating out $N$ and $S$, one obtains the usual 
low-energy supersymmetry potential with the parameters
$\mu$ and $\bmu$ given by
\bea
\mu &=&-\frac{5}{32\pi^2}\frac{\lambda\lambda_1}{\lambda_2}\frac{F^2}
{M_N^2M}~,
\label{zip}\\
\bmu &=&-\frac{10\lambda^2\lambda_1
\lambda_2}{(16\pi^2)^2}
\left( 1+\frac{5F^2}{8\lambda_2^2M_N^4}\right)\frac{F^2}{M^2}~.
\label{zap}
\eea
Thus, this model generates
$\bmu\sim\mu^2\sim m^2_0$, instead of
$\bmu\sim\mu F/M$. 

There is another way  to understand why this mechanism works, based on 
 a pseudo-Goldstone boson interpretation.
Let me modify the previous model by introducing a new gauge singlet
$\bar N$ and by replacing eq.~(\ref{spot}) by
\be
W=S(\lambda_1 H \bar H +{\lambda_2}N\bar N+\lambda \Phi \bar \Phi
-M_N^2) ~.
\ee
The results of the model (\ref{spot}) 
are essentially unaffected. However here,
in the limit $\lambda_1 = \lambda_2$, the superpotential 
has a U(3) symmetry under which $\Sigma\equiv (H,N)$ and
 $\bar{\Sigma}\equiv (\bar H, \bar N)$ 
transform as a triplet and an anti-triplet.
In the supersymmetric limit,
the VEVs of $N$ and $\bar N$ break the
U(3) spontaneously to U(2) and
the two Higgs doublets are identified with the corresponding 
Goldstone bosons.
Actually they are only pseudo-Goldstone
bosons since they get non-zero masses as soon as gauge 
and quark-Yukawa interactions
are switched-on.
Nevertheless,
at the one-loop level,
the relevant part of the effective potential is still U(3)-invariant
and  one  combination of the two Higgs doublets
$({H + \bar H^\dagger)/\sqrt{2}}$, remains exactly massless.
Indeed, at one loop, 
the determinant of the Higgs mass-squared matrix is zero, and
\be
\bmu=|\mu|^2\, ,\label{pgbeq}
\ee
a general property of models in which the Higgs particles are 
pseudo-Goldstone bosons \cite{fine}. Soft masses for $H$ and $\bar H$
are generated at two loops by the 
 gauge contribution eq.~(\ref{masses}). This latter
violate the U(3) invariance and then the determinant 
of the Higgs mass-squared matrix no longer vanishes.
Nevertheless, we are still guaranteed to obtain a $\mu$ and $\bmu$ of
the correct magnitude, since eq.~(\ref{pgbeq}) is spoiled only by
two-loop effects. If we now allow $\lambda_1 \ne \lambda_2$, we will
modify eq.~(\ref{pgbeq}) but not the property $\bmu \sim \mu^2$.
This provides an alternative explanation
of why the above mechanism can work.

Let me note that, surprisingly,
large values for $\mu$ and $\bmu$  ($\sim$ 1 TeV$\gg M_Z$)
do not always mean violations
of  naturalness. 
One could  think of a scenario in which 
the weak scale is  protected by the above U(3) symmetry.

\begin{table*}[htb]
\setlength{\tabcolsep}{1.5pc}
\newlength{\digitwidth} \settowidth{\digitwidth}{\rm 0}
\catcode`?=\active \def?{\kern\digitwidth}
\caption{\centerline{SUSY--GUT PROBLEMATICS}}
\label{tab:effluents}
\begin{tabular*}{\textwidth}{@{}l@{\extracolsep{\fill}}rrrr}
\hline
                 & \multicolumn{1}{r}{Gravity Mediated} 
                 & \multicolumn{1}{r}{Gauge Mediated} \\
\hline
Flavor problem    &$$  & $\star$\\
$\mu$-problem    &$\star$  &$$ \\
Doublet-triplet splitting problem & $$ & $\star$\\
Dark matter    problem&$\star$  & $$\\
\hline
\end{tabular*}
\end{table*}

\section{DOUBLET-TRIPLET SPLITTING\\ AND DARK MATTER PROBLEM}

The doublet-triplet splitting problem arises in GUT 
such as SU(5) where the Higgs doublet
is embedded in a GUT-representation with a color triplet; while
the Higgs doublet has to be light to break the electroweak symmetry
at low energies, the color triplet has to be heavy to avoid
a large proton decay or a mismatch of the gauge couplings at the GUT scale.
Several mechanisms have been proposed in the literature to generate
this mass splitting. The most economical one is the sliding 
singlet \cite{witten}:
An extra singlet is introduced in the 
theory  whose  VEV dynamically adjusts to  
produce the doublet-triplet mass splitting.
Nevertheless, this mechanism  cannot work in  theories
with supersymmetry broken at high energies 
\cite{sliding}.  This is because a tadpole of order $F^2/M_P$ 
is induced for the singlet, such that it
shifts  its VEV to a value where both doublet and triplet
are heavy. Clearly, for small values of the supersymmetry
breaking scale, $\sqrt{F}\sim$ 10 TeV, 
the above tadpole is not dangerous and the sliding-singlet mechanism 
can be operative.

Let me finally turn to the dark matter problem. It is well known
that GravMSB
theories
have a natural candidate for dark matter, the neutralino. 
This is usually 
the
  lightest supersymmetric particle (LSP). It is  then
 stable and can populate the present universe as a relic of the hot
primordial era.
For theories with
 supersymmetry  breaking 
at low energies, the LSP is the gravitino
that can only be a dark matter candidate for $\sqrt{F}\sim
10^3$ TeV \cite{gravitino}. This value results too large for  
GMSB theories with one scale, 
$M\sim\sqrt{F}\sim $10--100 TeV. 
Candidates for dark matter can be found, however, in the
 messenger or hidden sector   \cite{dgp}.

\section{CONCLUSIONS}

As we have seen, the two scenarios, GravMSB and GMSB,
face differently the above  supersymmetric problems.
Both suffer from some drawbacks.
Of course, 
these drawbacks can always
be overcome by complicating the models.
My personal point of view is summarized in table 1, where
for each problem a ``star'' 
is given to the best suited scenario.

\vspace{.2cm}
 
It is a pleasure to thank  my collaborators Savas Dimopoulos, Gia Dvali and
Gian Giudice 
 for numerous enlightening discussions that led to the work described above.

\end{document}